\documentclass[prl,twocolumn,showpacs,preprintnumbers,amsmath,amssymb,citeautoscript]{revtex4-1}

\usepackage{graphicx,tabularx}
\usepackage{amssymb, amsfonts, amsmath}
\usepackage{dcolumn}
\usepackage{bm}
\usepackage{color}

\emergencystretch=\maxdimen
\newlength{\figwidth}
\begin{document}

\title{{Fully gapped superconductivity with no sign change  in the prototypical  heavy-fermion  CeCu$\boldsymbol{_2}$Si$\boldsymbol{_2}$
}}

\author{T.\;Yamashita$^{1,*}$}
\author{T.\;Takenaka$^{2,*}$}
\author{Y.\;Tokiwa$^{1,*}$}
\author{J.\,A.\;Wilcox$^{3,*}$}
\author{Y.\;Mizukami$^2$}
\author{D.\;Terazawa$^1$}
\author{Y.\;Kasahara$^1$}
\author{S.\;Kittaka$^4$}
\author{T.\;Sakakibara$^4$}
\author{M.\;Konczykowski$^5$}
\author{S.\;Seiro$^6$}
\author{H.\,S.\;Jeevan$^6$}
\author{C.\;Geibel$^6$}
\author{C.\;Putzke$^3$}
\author{T.\;Onishi$^1$}
\author{H.\;Ikeda$^7$}
\author{A.\;Carrington$^{3,\dagger}$}
\author{T.\;Shibauchi$^{2,\dagger}$}
\author{Y.\;Matsuda$^{1,\dagger}$}

\medskip
\affiliation{
	$^1$Department of Physics, Kyoto University, Kyoto 606-8502, Japan\\
	$^2$Department of Advanced Materials Science, University of Tokyo, Kashiwa, Chiba 277-8561, Japan \\
	$^3$H.\,H.\,Wills Physics Laboratory, University of Bristol, Bristol BS8 1TL, UK\\
	$^4$Institute for Solid State Physics, University of Tokyo, Kashiwa, Chiba 277-8581, Japan \\
	$^5$Laboratoire des Solides Irradi{\'e}s, {\'E}cole Polytechnique, CNRS, CEA, Universit{\'e} Paris-Saclay, 91128 Palaiseau Cedex, France \\
	$^6$Max Planck Institute for Chemical Physics of Solids, N\"othnitzer Strasse 40, 01187 Dresden, Germany\\
	$^7$Department of Physics, Ritsumeikan University, Kusatsu 525-8577, Japan\\
	$^*$These authors contributed equally to this work.\\
	$^\dagger$Corresponding authors. Email: {\sf A.Carrington@bristol.ac.uk} (A.C.); {\sf shibauchi@k.u-tokyo.ac.jp} (T.S.); {\sf matsuda@scphys.kyoto-u.ac.jp} (Y.M.)
}
\date{\today}

\begin{abstract}{\bf
In exotic superconductors including high-$\boldsymbol{T_c}$ copper-oxides, the interactions mediating electron Cooper-pairing are widely considered to have a magnetic rather than the conventional electron-phonon origin. Interest in such exotic pairing was initiated by the 1979 discovery of heavy-fermion superconductivity in CeCu$\boldsymbol{_2}$Si$\boldsymbol{_2}$, which exhibits strong antiferromagnetic fluctuations. A hallmark of unconventional pairing by anisotropic repulsive interactions is that the superconducting energy gap changes sign as a function of the electron momentum, often leading to nodes where the gap goes to zero. Here, we report low-temperature specific heat, thermal conductivity and magnetic penetration depth measurements in CeCu$\boldsymbol{_2}$Si$\boldsymbol{_2}$, demonstrating the absence of gap nodes at any point on the Fermi surface. Moreover, electron-irradiation experiments reveal {that the superconductivity survives even when the electron mean free path becomes substantially shorter than the superconducting coherence length. This indicates that superconductivity is robust against impurities, implying that there is no sign change in the gap function. These results show} that, contrary to long-standing belief, heavy electrons with extremely strong Coulomb repulsions can condense into a fully-gapped $\boldsymbol{s}$-wave superconducting state, which has an on-site attractive pairing interaction.
}



\end{abstract}

\maketitle

\section*{Introduction}
The discovery of heavy-fermion superconductivity in CeCu$_2$Si$_2$  was an important turning point in the history of superconductivity, because it led to the birth of research on non-electron-phonon mediated pairing \cite{Steglich,Pfl}.  Heavy-fermion superconductivity is {usually} intimately related to magnetism in some form. In particular,  superconductivity often occurs in the vicinity of a zero-temperature magnetic instability known as a quantum-critical point (QCP) \cite{Mathur,Pfl,Thalmeier_rev}.  Thus it is widely believed that in these materials Cooper pairing is mediated by magnetic fluctuations.  The superconducting gap structure is a direct consequence of the mechanism producing the pairing.  In phonon-mediated conventional superconductors with a finite on-site pairing amplitude in real space (Fig.\,1A),  the superconducting gap function $\Delta(\bm{k})$ is  isotropic in momentum space (Fig.\,1B).    On the other hand, in magnetically-mediated unconventional superconductors,  the on-site pairing amplitude vanishes due to strong Coulomb repulsion and superconductivity is caused by a potential that is only attractive for particular displacements between the electrons forming the Cooper pair \cite{Scalapino} (Fig.\,1C).  A net attractive interaction can be realized if the superconducting gap changes sign on the Fermi surface (Fig.\,1D and E). In some materials, such as cuprates~\cite{Tsuei} and the heavy-fermion CeCoIn$_5$, the sign change of the gap leads to gap functions with nodes along certain momentum directions \cite{Tsuei,Yaz,Izawa}. However, in certain iron-pnictide superconductors the gap function has no nodes but may change sign between the well separated electron and hole Fermi surface pockets \cite{Mazin,Hirschfeld}.

\begin{figure}[t]
	\begin{center}
		
		\includegraphics[width=1.0\linewidth]{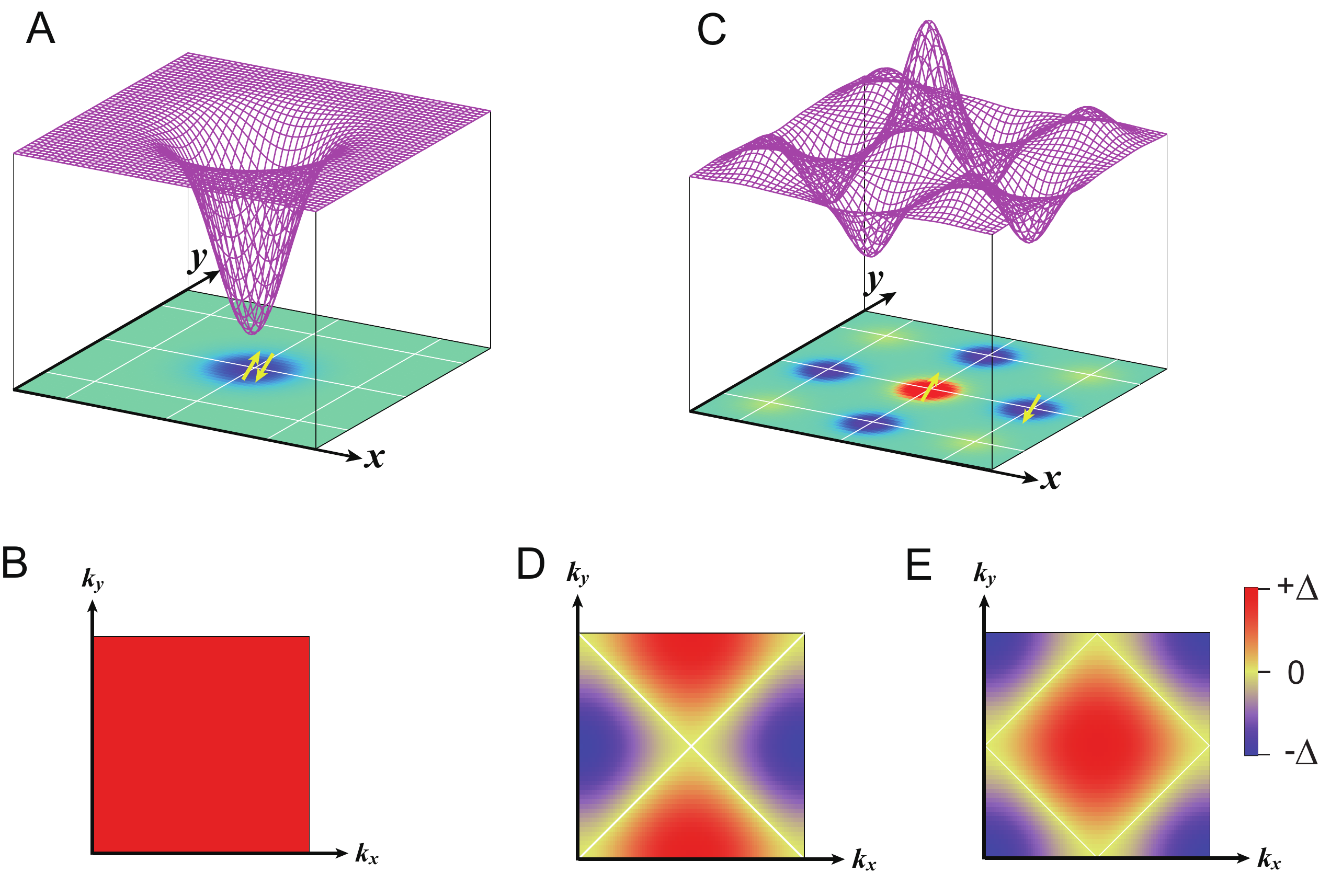}
		\caption{{\bf Pairing interactions and superconducting gap functions.}  ({\bf A}) The pairing interaction in real space for  attractive force mediated by electron-phonon interaction.  Blue part corresponds to attractive region. Both electrons composing the Cooper pair can occupy the same atom. ({\bf B}) Isotropic $s$-wave superconducting state in the momentum space driven by the attractive force shown in ({\bf A}). The gap function is constant  in the entire Brillouin zone. ({\bf C}) The pairing interaction due to  magnetic fluctuations.   The red and blue parts correspond to repulsive and attractive regions, respectively.   Both electrons cannot occupy the same atom.   Superconductivity is caused by the attractive part of the oscillating pairing interaction.  ({\bf D} and {\bf E}) Examples for the gap structures in momentum space for unconventional superconductors caused by an on-site repulsive force, $d_{x^2-y^2}$ symmetry ({\bf D}) and $s_{\pm}$ -symmetry ({\bf E}).   Due to the sign change of the superconducting order parameter, the gap vanishes  on the yellow lines.  When the Fermi surface crosses these lines, gap nodes appear.
		}
	\end{center}
	\vspace{-5mm}
\end{figure}

CeCu$_2$Si$_2$ is a prototypical heavy-fermion superconductor near a magnetic instability~\cite{Steglich,Yuan} with transition temperature $T_c \simeq 0.6$\,K~\cite{Steglich} (Fig.\,2A).  The Fermi surface consists of heavy electron and light hole bands (Fig.\,2B) \cite{Ikeda}.    Slight variations in stoichiometry lead to `$A$' type and `$S$' type crystals; the former is antiferromagnetic  and the latter is superconducting without magnetic ordering but lying very close to a magnetic QCP (Fig.\,2A). The in-plane resistivity above $T_c$ in zero field, which follows a power-law $\rho_a=\rho_{a0}+AT^{\epsilon}$ with $\epsilon=1.5$ (Fig.\,2C, inset) along with the heat capacity, which follows $C/T=\gamma_N-a\sqrt{T}$ in the normal state slightly above the upper critical field, are consistent with non-Fermi-liquid behaviors expected for three dimensional antiferromagnetic quantum critical fluctuations~\cite{Gegenwart,Zhu,Rosch}.
The magnetic-field-induced recovery of Fermi-liquid behavior with $\epsilon=2$ shown in Fig.\,2C bears striking resemblance to other heavy-fermion compounds in the vicinity of QCPs, such as CeCoIn$_5$ and YbRh$_2$Si$_2$~\cite{Paglione,Custers}. A critical slowing down of the magnetic response revealed by neutron scattering~\cite{Arndt} {and nuclear quadrupole resonance (NQR)~\cite{Ishida}  in the normal state,  has also been attributed to antiferromagnetic fluctuations near the QCP.}  These results have led to a wide belief that antiferromagnetic fluctuations are responsible for the pairing interaction in CeCu$_2$Si$_2$.
Here, we report a comprehensive study of the gap structure of $S$-type CeCu$_2$Si$_2$ using several different probes which together are sensitive to the gap structure on all Fermi surface sheets and also any possible changing of gap-sign between sheets.

\begin{figure}[t]
	\begin{center}
		\includegraphics[width=0.8\linewidth]{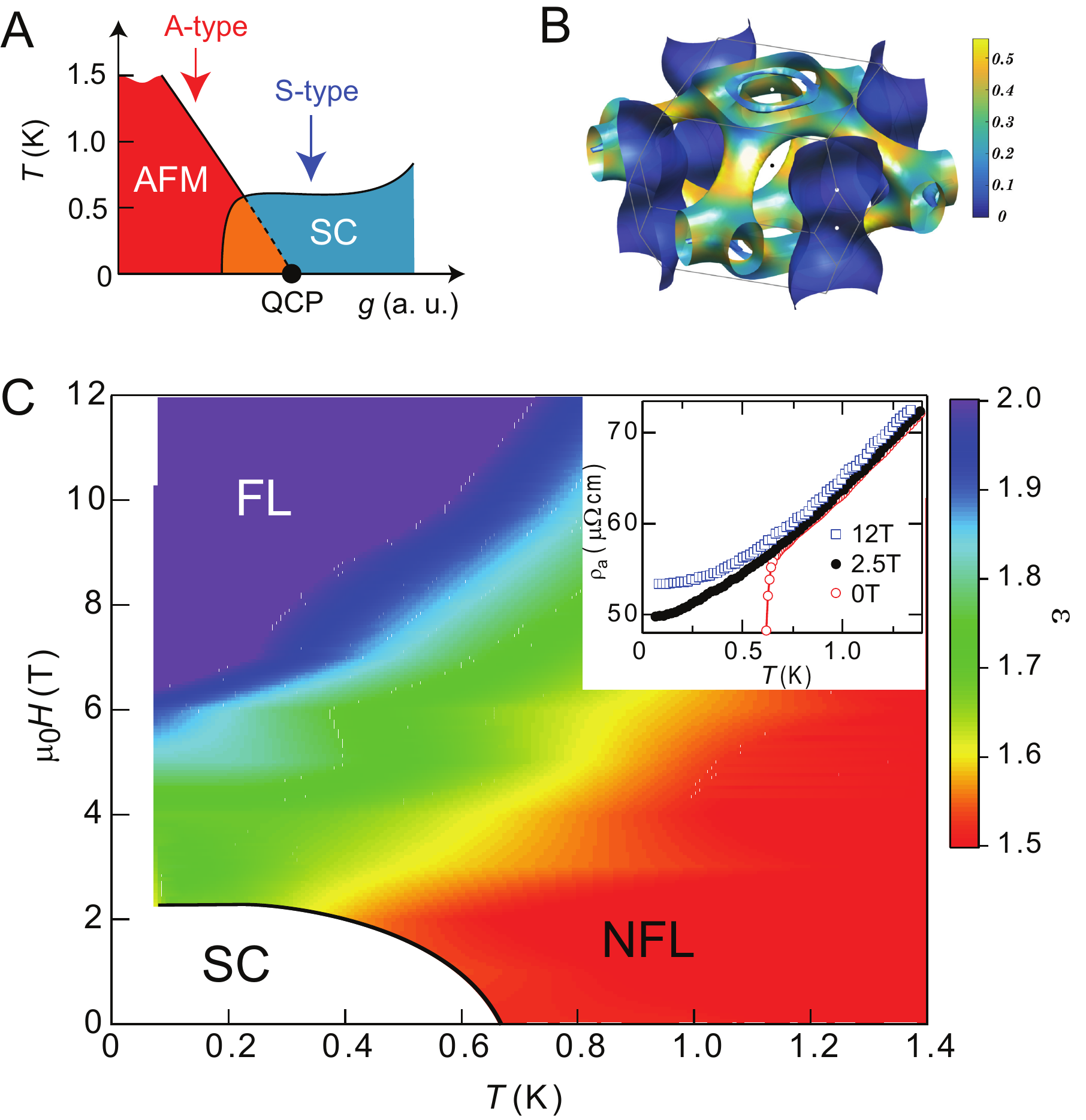}
		\caption{{\bf Phase diagrams and electronic structure of CeCu$\boldsymbol{_2}$Si$\boldsymbol{_2}$.}  ({\bf A}) Schematic $T$-$g$ phase diagram, where $g$ is a non-thermal control parameter, such as pressure, substitution or Cu-deficiency. Red and blue arrows indicate two different types of CeCu$_2$Si$_2$ with antiferromagnetic ($A$-type) and superconducting ($S$-type) ground states, respectively. $S$-type crystal locates very close to AFM QCP.   ({\bf B})  Fermi surface colored by the Fermi velocity (in units of $10^6$\,m/s) obtained by the LDA+U calculation \cite{Ikeda}.   Fermi surface consists of separated electron and hole pockets: heavy electron pockets with cylindrical shape around $X$-point and rather complicated light hole pockets centred at $\Gamma$-point. ({\bf C}) $H$-$T$ phase diagram with color-coding of $T$-exponent ($\epsilon$) of the in-plane electrical resistivity, $\rho(T)=\rho_0+AT^{\epsilon}$  for $\bm{H}\parallel c$.  Inset shows the $T$-dependence of  $\rho(T)$ in zero field and in magnetic fields of 2.5 and 12\,T applied along the $c$ axis.
		}
	\end{center}
	\vspace{-5mm}
\end{figure}

\section*{Results}

\subsection*{Specific heat}

\begin{figure}[t]
	\begin{center}
		\includegraphics[width=1.0\linewidth]{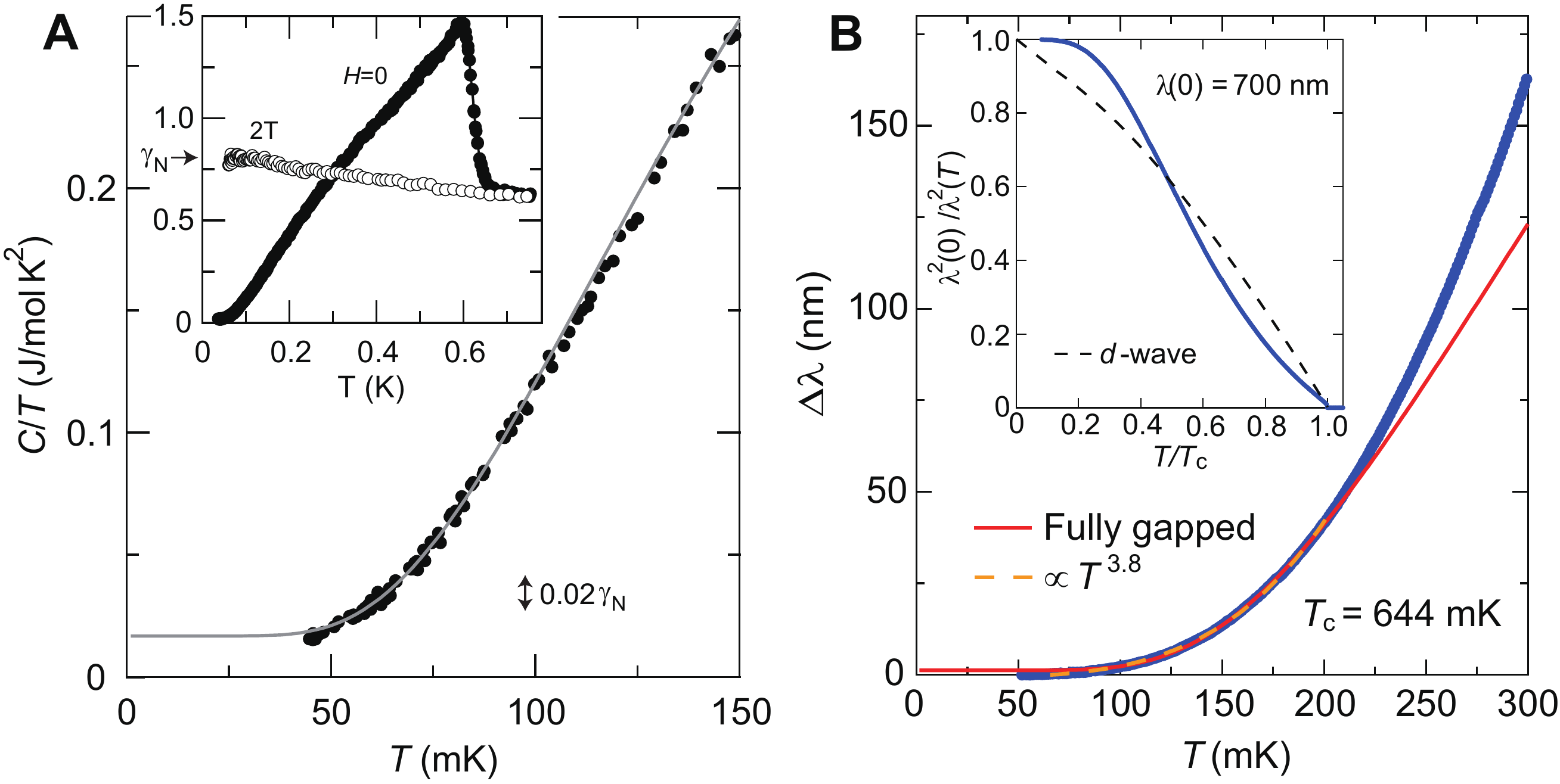}
		\caption{{\bf Temperature dependencies of specific heat and London penetration depth well below the superconducting transition temperature $\boldsymbol{T_c}$.} ({\bf A}) Inset shows the specific heat divided by temperature $C/T$ in zero field and in the normal state at $\mu_0H=2$\,T  for {\boldmath $H$}$\parallel ab$ plane.  The main panel shows $C/T$ at low temperatures. The gray solid line is an exponential fit of the data, yielding $\Delta=0.39$\,K.  ({\bf B}) Temperature dependent change in the in-plane penetration depth $\Delta\lambda$ in a single crystal of CeCu$_2$Si$_2$.	 The dashed (solid) line is a fit to a power-law (exponential) temperature dependence up to 0.2\,K. Inset shows the normalized superfluid density $\rho_s(T)=\lambda^2(0)/\lambda^2(T)$ as a function of $T/T_c$,	extracted by using a value of $\lambda(0)=700$\,nm (see section SIII in the Supplementary Materials). The line is the temperature dependence of $\rho_s(T)$ in the simple $d$-wave case. }
	\end{center}
	\vspace{-5mm}
\end{figure}

Specific heat $C$ is a bulk probe which measures {all thermally induced excitations.} Figure\,3A and its inset depict the specific heat $C/T$ for a crystal used in the present study.
At zero field $C/T$ exhibits a sharp transition at $T_c$ and tends towards saturation at the lowest temperature. The $C/T$ value at the lowest temperature, 15\,mJ/K$^2$mol, is less than 2\% of $\gamma_N$, which is nearly half of that in the previous report \cite{Kittaka}.
This extremely small $C/T$ indicates a very low number of quasiparticle excitations and that any inclusion of non-superconducting `$A$' type material is very small.
The data are well fitted by an exponential $T$ dependence showing a lack of thermally induced excitations at the lowest temperatures in agreement with {previous studies \cite{Kittaka,KittakaPRB}.} A linear behavior does not fit our $C/T$ data but if it was forced to then a fit above 90\,mK in Fig.\,3A would lead to an unphysical negative intercept at $T=0$\,K.  This is indicative of a fully gapped state with minimal disorder.  {More precisely, since the specific heat is dominated by the parts of the Fermi surface where the Fermi velocity is low (or mass large),} the $C/T$ data suggest the absence of line nodes in the \textit{heavy} electron band.

\subsection*{Penetration depth and lower critical field}
The magnetic penetration depth by contrast, measures the surface of the sample (to depth of a few micron), and is dominated by the low mass, high velocity parts of the Fermi surface.  We find that the in-plane penetration depth $\lambda_{ab}(T)$  at low temperatures ($T \ll T_c$) exhibits strong curvature and tends towards becoming $T$-independent (Fig.\,3B), similar to the results for $C/T$ and in contrast to the $T$-linear dependence expected for clean superconductors with line nodes~\cite{Bonn}.   A fit to a power-law $T$ dependence  $\Delta\lambda(T)(=\lambda_{ab}(T)-\lambda_{ab}(0))\propto T^n$ gives a high power $n>3.5$ (see section SI and Figs.\,S1, S2 in the Supplementary
Materials), which is practically indistinguishable from the exponential dependence expected in fully gapped superconductors.  {Since $\lambda_{ab}$ measures the in-plane superfluid response, our data show that gap nodes, at which quasiparticles with momentum  parallel to the $ab$ plane are excited, are absent on the light hole bands.}

{For more detailed analysis of the superconducting gap structure, the absolute value of $\lambda_{ab}(0)$ is necessary so that the  normalized superfluid density $\rho_s(T)=\lambda_{ab}^2(0)/\lambda_{ab}^2(T)$ can be calculated.   Unfortunately previous measurements have reported a wide spread of values of $\lambda_{ab}(0)$ (120 to 950\,nm \cite{Pene1,Pene2}) which probably reflects differences in sample stoichiometry between studies.  We have estimated  $\lambda_{ab}(0)=700$\,nm  from Hall-probe magnetometery measurements of the lower critical field $H_{c1}$ in the same samples as used for our $\Delta\lambda(T)$ study (see section SII and Fig.\,S3 in the Supplementary Materials).  The inset of Fig.\,3B shows the $T$-dependence of  $\rho_s(T)$. Near $T_c$, we find convex curvature in $\rho_s(T)$, which is a signature frequently observed in multigap superconductors \cite{Prozorov_rev}.} {Indeed, the two-gap behavior has been reported in the recent scanning tunneling spectroscopy \cite{Enayat} and specific heat measurements \cite{Kittaka,KittakaPRB}. }

\begin{figure}[t]
	\begin{center}
		\includegraphics[width=1.0\linewidth]{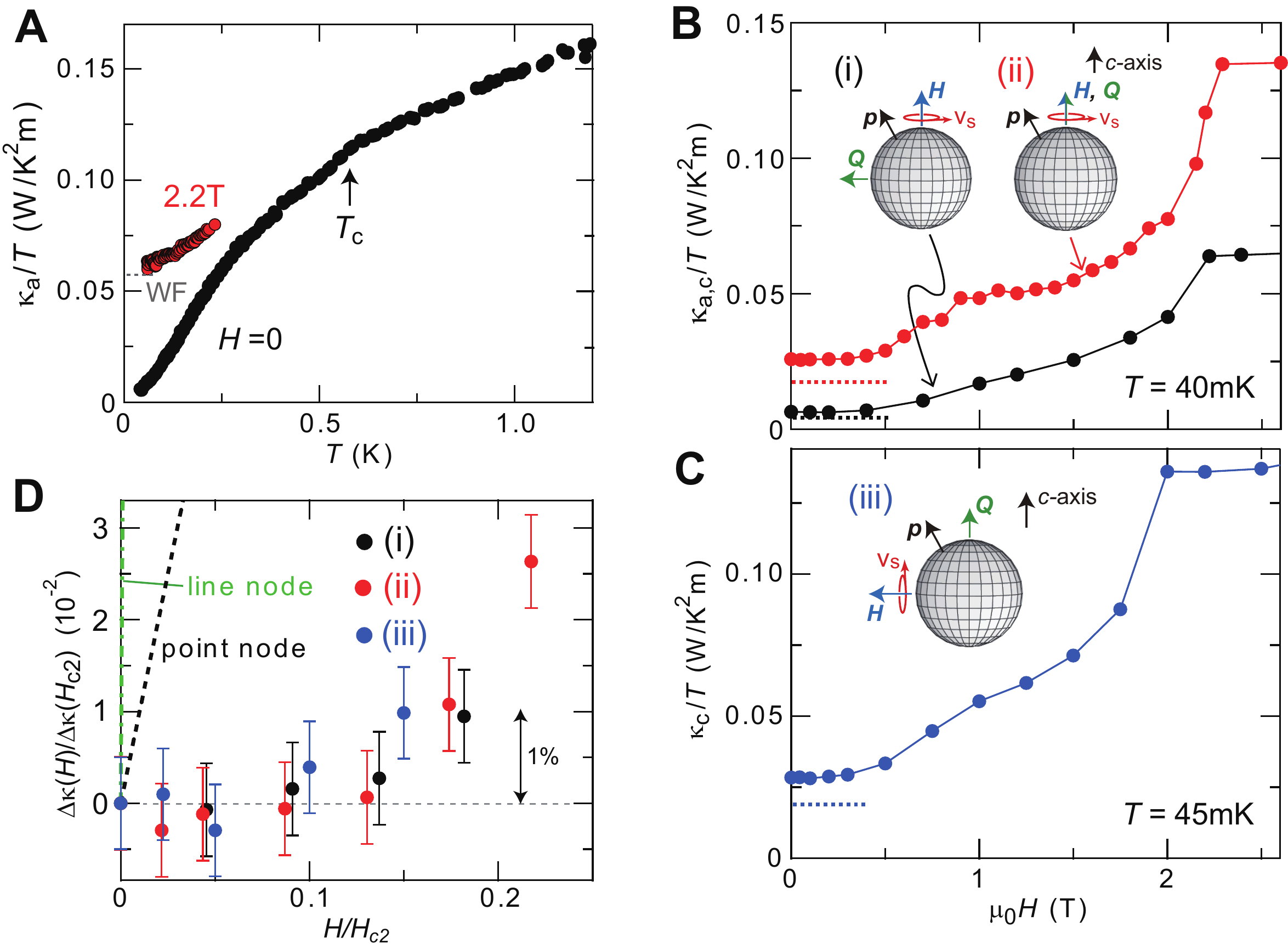}
		\caption{{\bf Thermal conductivity of CeCu$\boldsymbol{_2}$Si$\boldsymbol{_2}$ for various directions of thermal current and magnetic field.}  ({\bf A}) Temperature dependence of the in-plane thermal conductivity divided by temperature $\kappa_{a}/T$ in zero field and in magnetic field of $\mu_0H=2.2$\,T applied along the $c$ axis.  WF refers to $\kappa/T$ at $T\rightarrow 0$ calculated from the Wiedemann-Franz law. ({\bf B}) Field dependence of $\kappa/T$ for two different configurations.  (i) $\kappa_{a}/T$ ($\bm{Q}\parallel a$) in $\bm{H}\parallel c$ and (ii) $\kappa_c/T$ ($\bm{Q}\parallel c$) in  $\bm{H}\parallel c$.  In these configurations, thermal conductivity selectively probes the excited quasiparticles with in-plane momentum.  The dashed horizontal lines represent the phonon contribution, $\kappa_{ph}/T$,  estimated from the WF law above upper critical field (see the main text). ({\bf C}) Field dependence of $\kappa_c/T$ for configuration (iii), where $\bm{Q}\parallel c$ and $\bm{H}\parallel a$.  In this case, thermal conductivity selectively probes the excited quasiparticles with out-of-plane momentum. ({\bf D}) Field-induced enhancement of thermal conductivity $\Delta\kappa(H)\equiv \kappa(H)-\kappa(0)$ normalized by the normal state value, $\Delta\kappa(H)/\Delta\kappa(H_{c2})$ for the configurations (i), (ii) and (ii) plotted against the magnetic field normalized by the upper critical fields.  Black and green broken lines represent the field dependencies expected for line and point nodes.   }
	\end{center}
	\vspace{-5mm}
\end{figure}

\subsection*{Thermal conductivity}
Thermal conductivity is a bulk, directional probe of the quasiparticle excitations, and like penetration depth, is dominated by the high velocity parts of the Fermi surface \cite{Matsuda}.  Figure\,4A and its inset show the $T$-dependence of the in-plane thermal conductivity  $\kappa_{a}/T$  (with heat current $\bm{Q}\parallel a$).  The thermal conductivity in the normal state at $T\rightarrow 0$  slightly above the upper critical field for $\bm{H}\parallel c$ obeys well the Wiedemann-Franz law, $\kappa_a/T=L_0/\rho_a$ (Fig.\,4A, dashed line), where $L_0$  is the Lorenz number and $\rho_a$ is the in-plane resistivity.  At the lowest temperatures $\kappa_a/T$ extrapolated to $T=0$ is zero within our experimental resolution and is at least an order of magnitude smaller than that expected for line nodes (see section SIII and Fig.\,S4 in the Supplementary Materials), consistent with the  $\Delta\lambda(T)$ results.

Further evidence for the absence of any nodes is provided by $H$-dependence of $\kappa$. In fully gapped superconductors,  where all the quasiparticles states are bound to vortex cores, the magnetic field hardly affects $\kappa$ except in the vicinity of the upper critical field $H_{c2}$.  By contrast, in  nodal superconductors, heat transport is dominated by the delocalized quasiparticles.  In the presence of a supercurrent with velocity ${\bm v}_s$ around the vortices, the energy of a quasiparticle with momentum $\bm{p}$ is Doppler shifted relative to the superfluid by  $E({\bm p})\rightarrow E({\bm p})-{\bm v}_s\cdot {\bm p}$, giving rise to an initial steep increase of $\kappa(H)/T\propto \sqrt{H}$ for line nodes and $\kappa(H)/T\propto H \log H$ for point nodes. Thermal conductivity selectively probes the quasiparticles with  momentum parallel to the thermal current ($\bm{p}\cdot\bm{Q}\neq 0$) and with momentum perpendicular to the magnetic field ($\bm{p}\times\bm{H}\neq 0$) because $\bm{H}\perp\bm{v}_s$ \cite{Matsuda}.   To probe the quasiparticle excitations on the whole Fermi surface, we performed measurements for three different configurations,  (i) $\kappa_a$ for $\bm{H}\parallel c$, (ii) $\kappa_c$ for $\bm{H}\parallel c$, and (iii) $\kappa_c$ for $\bm{H}\parallel a$ (Figs.\,4B, C).   For (i) and (ii), thermal conductivity selectively probes the quasiparticles with in-plane momentum,  whereas for (iii) it selectively probes quasiparticles with out-of-plane momentum. For configuration (ii), there is structure at  $\mu_0H \sim1$\,T,  which again indicates the presence of multiple superconducting gaps.  The $H$-dependence for configuration (iii) shown in Fig.\,4D is similar to configuration (i).    Remarkably, in all configurations, magnetic field hardly affects the thermal conduction in the low field regime (Figs.\,4B, C);  the field-induced enhancement, $\Delta\kappa(H)\equiv \kappa(H)-\kappa(0)$  is less than 1/100 of the normal-state value  $\Delta\kappa(H_{c2})$ even at $H/H_{c2}\sim0.15$,  demonstrating  a vanishingly small number of delocalized quasiparticles  excited by  magnetic field.    As shown by the dashed lines in  Fig.\,4D, $\Delta\kappa(H)/\Delta\kappa(H_{c2})$ is far smaller than that expected for line and point nodes.

\begin{figure}[t]
	\begin{center}
		\includegraphics[width=0.9\linewidth]{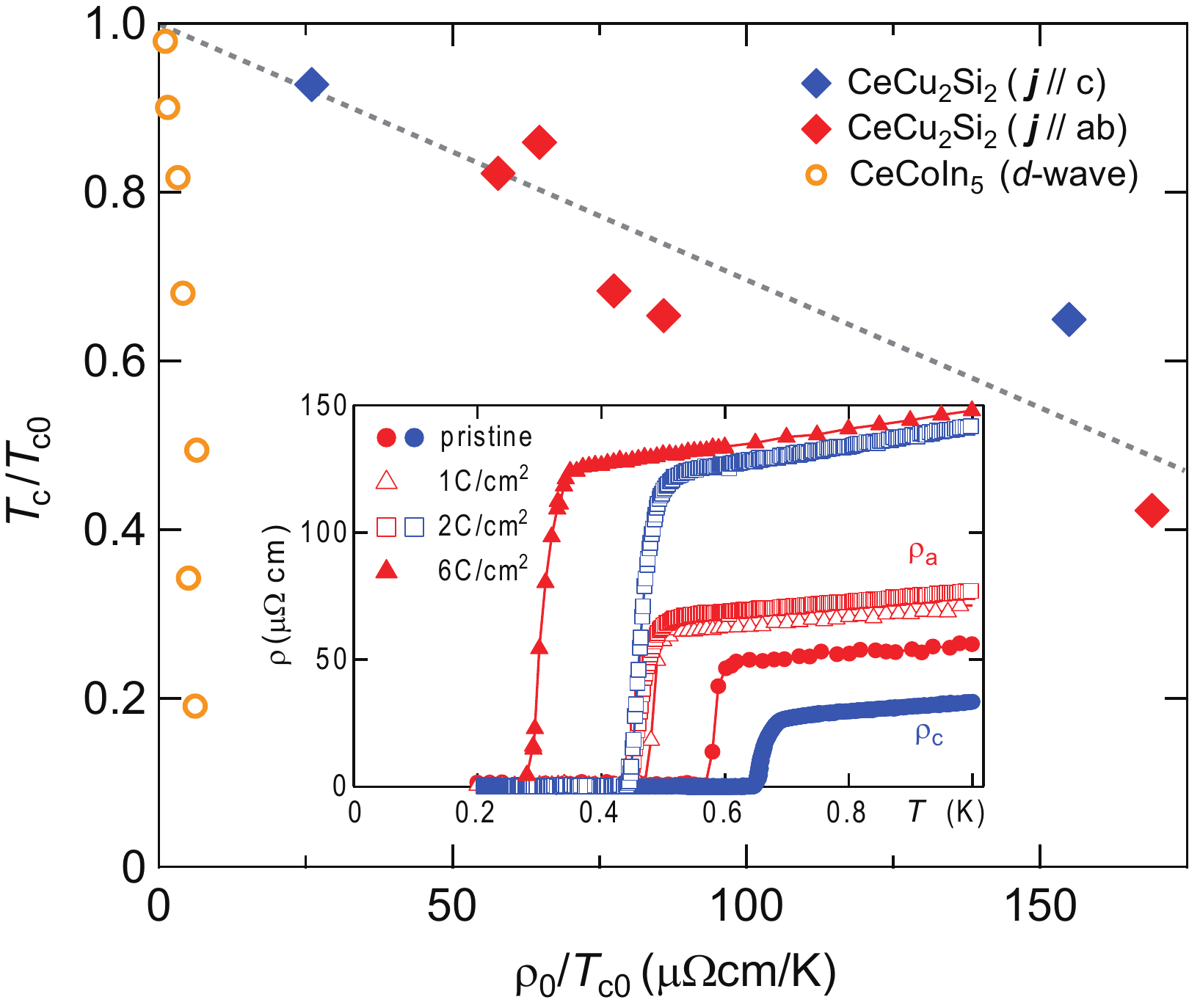}
		\caption{{\bf Pair-breaking effect of CeCu$\boldsymbol{_2}$Si$\boldsymbol{_2}$. }    Suppression of superconducting transition temperature $T_c/T_{c0}$ as a function of $\rho_0/T_{c0}$, which is proportional to the pair breaking parameter,  for CeCu$_2$Si$_2$ and Sn-substituted CeCoIn$_5$ ($d$-wave)~\cite{Bauer}.  Here $T_{c0}$ is the transition temperature with no pair-breaking effect and $\rho_0$ is the residual resistivity.   For CeCu$_2$Si$_2$,  $T_{c0}=0.71$\,K is used.   Inset shows the temperature dependence of resistivity in  CeCu$_2$Si$_2$ before and after electron irradiation that creates point defects.
		}
	\end{center}
	\vspace{-5mm}
\end{figure}

\subsection*{{Electron irradiation}}
{The above measurements of $C(T)$, $\Delta\lambda(T)$ and $\kappa(T,H)$ demonstrate the absence of any kind of nodes in the gap function on the whole Fermi surface. To further distinguish between the remaining possible gap structures we have measured the effect of impurity-induced pair-breaking on $T_c$.  These measurements are a sensitive test of possible sign changes in the gap function either between different Fermi surface sheets or on a single sheet.  Impurity induced scattering between sign changing areas of Fermi surface will reduce $T_c$ very rapidly whereas if there is no sign change the reduction will be much slower or even zero. To introduce impurity scattering by homogeneous point defects in a controllable way, we employed electron irradiation with incident energy of 2.5\,MeV \cite{Mizukami}, which according to our calculation of electron scattering cross sections, mainly removes Ce atoms.   Electronic-structure calculations of CeCu$_2$Si$_2$ \cite{Ikeda} show that the bands crossing the Fermi level are mainly composed of a single Ce $f$-manifold, so removing Ce atoms by electron irradiation will act as a strong point scatterer and induce both intra- and inter-band impurity scattering with similar amplitude.}

{Our results show that $T_c$ of CeCu$_2$Si$_2$ is decreased slowly} with increasing dose (inset of Fig.\,5).   The transition width remains almost unchanged after irradiation, indicating good homogeneity of the point defects. {The temperature dependence of resistivity indicates that the primary effect of irradiation is the increase of temperature-independent impurity scattering with dose, and that the temperature-dependent inelastic scattering remains unaffected.} In- and out-of-plane residual resistivities reach $\rho_{a0}\sim120\,\mu\Omega$cm and $\rho_{c0}\sim110\,\mu\Omega$cm for irradiated crystals (inset of Fig.\,5). Using $\ell_{j}=v_F^{j}\lambda_{j}^2(0)\mu_0/\rho_{j0}$ ($j=ab$ or $c$), we obtain in- and out-of-plane mean free paths, $\ell_{ab}\sim3.0$\,nm and $\ell_c\sim 1.8$\,nm, respectively. Here, we used averaged in-plane (out-of-plane) Fermi velocity $v_F^{ab}\sim5800$\,m/s ($v_F^c\sim6800$\,m/s) of the light hole band, $\lambda_c(0)=\lambda_{ab}(0)(\xi_{ab}/\xi_c)=480$\,nm, where in-plane and out-of-plane coherence lengths determined by the orbital limited upper critical fields, $\xi_{ab}=4.7$\,nm and $\xi_c=6.9$\,nm, respectively (see section SIV in the Supplementary Materials). These mean free paths are obviously shorter than $\xi_{ab}$ and $\xi_c$.  For unconventional pairing symmetries such as $d$-wave, superconductivity is completely suppressed at $\ell \lesssim 4 \xi$.  {In stark contrast,  $T_c$ of CeCu$_2$Si$_2$ is still as high as $\sim T_{c0}/2$ even for $\ell_c/\xi_c\sim0.26$ and $\ell_{ab}/\xi_{ab}\sim0.64$. We note that this $\ell/\xi$ is the upper limit value, because $\ell$ is estimated from the penetration depth and conductivity, both of which are governed by the light bands, while $\xi$ is determined by the upper critical field which is governed by heavy bands. Thus this result demonstrates that superconductivity in CeCu$_2$Si$_2$ is robust against impurities. This is also} seen clearly by comparison to the $d$-wave superconductor CeCoIn$_5$ \cite{Bauer}, which has comparable effective mass and carrier number. Figure\,5 displays the residual resistivity dependence of $T_c/T_{c0}$, where $T_{c0}$ is the  transition temperature with no pair breaking.   In CeCoIn$_5$, $T_c$  is suppressed to zero in the sample with $\rho_0/T_{c0}$ smaller than 10\,$\mu\Omega$cm/K~\cite{Bauer}, while $T_c$  in CeCu$_2$Si$_2$ is still $\sim50$\,\% of $T_{c0}$ even for the sample with $\rho_0/T_{c0}$  larger than 150\,$\mu\Omega$cm/K, indicating that the pair breaking effect in CeCu$_2$Si$_2$ is fundamentally different from that in CeCoIn$_5$.  

{Comparison to other materials (cuprates and iron-pnictides) with sign changing gaps confirms the much weaker effect of impurities in CeCu$_2$Si$_2$.  In Fig.\,S5 we plot the impurity induced $T_c$ reduction in a number of materials as a function of the scattering rate, estimated from $\rho_0/\lambda$, divided by $T_{c0}$. Plotting the data in this way takes out the effect of difference in $T_c$ and carrier density between the different materials, and it can be seen that the $T_c$ reduction in CeCu$_2$Si$_2$ is much weaker than the archetypal cuprate YBa$_2$Cu$_3$O$_7$ \cite{Rullier} which has a sign-changing $d_{x^2-y^2}$ gap function.  The iron-pnictides present an very unusual system where the $\bm{k}$-dependence of the scattering is critical to the effect of impurities on $T_c$.  Assuming that the pairing in these materials is caused by inter-band spin-fluctuation interactions, then the gap function will change sign between the electron and hole sheets ($s_\pm$ pairing). Inter-band impurity scattering will then increase $\rho_0$ and decrease $T_c$ in a similar way to other sign-changing gap materials, however, if the scattering is purely intra-band then this would increase $\rho_0$ but would not decrease $T_c$ \cite{Wang}.   It is highly unlikely that such an anomalous situation could occur in CeCu$_2$Si$_2$ because the Fermi surface sheets are not well separated and, as described above, Ce vacancies would produce non-$\bm{k}$-selective scattering.}

{The slow but finite reduction in $T_c$ as function of $\rho_0$ we see in CeCu$_2$Si$_2$ can be explained qualitatively by the moderate gap anisotropy we have observed in our $C(T)$ and $\lambda(T)$ measurements. In cases where there is gap anisotropy, scattering will tend to average out the gap thus depressing $T_c$. However, crucially this will be at a much slower rate than for a sign changing gap, as illustrated in Fig.\ S5 by data for the non-sign changing $s$-wave superconductors MgB$_2$ \cite{Putti} and YNi$_2$B$_2$C \cite{Karkin} which are known to have very anisotropic energy gaps.  Our observed slower decrease in $T_c$ as a function of impurity scattering in CeCu$_2$Si$_2$ compared to these materials is consistent with our observed moderate anisotropy.}

\section*{Discussion}

{The combination of our measurements and previous results rules out all but one possible gap structure.  The strong reduction of the spin susceptibility in the superconducting state observed by nuclear magnetic resonance Knight shift indicates spin singlet pairing \cite{Kitaoka} which rules out any odd-momentum ($p$ or $f$) states, including those, such as the Balain-Werthamer state \cite{Balian} which are fully gapped \cite{Sigrist}. This is consistent with the observation that $H_{c2}$ is Pauli limited \cite{Vieyra}.  In fact, in the present crystal, orbital-limited upper critical fields at $T=0$ calculated from $H_{c2}^{\rm orb}=-0.7T_c(dH_{c2}/dT)_{T_c}$ are 10.0 and 14.7\,T for $\bm{H}\parallel a$ and $\bm{H}\parallel c$, respectively.  These values are much larger than the observed $H_{c2}$ of 2.0\,T for $\bm{H}\parallel a$ and 2.3\,T for $\bm{H}\parallel c$.  }

{Our observation that superconductivity is robust against inter-band and intra-band impurity scattering rules out any sign-changing gap functions such as $d$-wave or the recently proposed sign changing $s_{\pm}$ state \cite{Ikeda}.  Both the $d$-wave and $s_{\pm}$ states are also highly unlikely because neither could be nodeless in CeCu$_2$Si$_2$ where the electron and hole Fermi surface sheets are not well separated.  Finally, unconventional states which combine irreducible representations of the gap function, such as $d_{xy}+id_{x^2-y^2}$ or $s+id_{x^2-y^2}$ can be ruled out because such states would be highly sensitive to impurities and furthermore as these representations are not in general degenerate we would expect to see two distinct superconducting transitions.  If there was accidental degeneracy, this would be broken by pressure or doping but no double transitions are observed in these conditions either \cite{Yuan}.  This leads us to the surprising conclusion that the pairing in CeCu$_2$Si$_2$ is a fully-gapped non-sign changing $s$-wave state.}

Previously, evidence for line-nodes in CeCu$_2$Si$_2$ has been suggested by measurements of the NQR relaxation rate $1/T_1$ where a $T^3$-dependence  below $T_c$ was observed \cite{Ishida}.  However, these results would also be explained by the multigap nature of the superconductivity shown here by our $C(T)$, $\lambda(T)$ and $\kappa(H)$ measurements.  {The absence of the coherence (Hebel-Slichter) peak in $1/T_1(T)$ below $T_c$ may be explained by the quasiparticle damping especially for anisotropic gap and thus does not give conclusive evidence for the sign changing gap \cite{KittakaPRB}.} Inelastic neutron scattering shows an enhancement of magnetic spectral weight at around $E\sim 2\Delta$ \cite{Stockert} which could be interpreted in terms of a spin-resonance expected in superconductors with a sign-changing gap.  However, this enhancement is very broad compared with some cuprates \cite{Keimer} and CeCoIn$_5$ \cite{Stock} so is not clearly a resonance peak which is expected to be sharp in energy. Moreover, recent calculations show that a broad maximum at $E\sim 2\Delta$ appears even in superconductors without sign changing gaps \cite{Onari}. {We should add that even in the $d$-wave CeCoIn$_5$ case, the interpretation of the neutron peak below $T_c$ is still controversial \cite{Chubukov,Song}.} Hence the NQR and neutron results do not provide conclusive evidence for a sign changing gap structure and are not necessarily inconsistent with the results here.

{At first sight our finding that CeCu$_2$Si$_2$ has a non-sign changing $s$-wave gap function casts doubt on the long-standing belief that it is a magnetically-driven superconductor, despite overwhelming evidence that this compound is located near a magnetic QCP.  It is unlikely that the conventional electron-phonon interaction could overcome the on-site strong Coulomb repulsive force, which enhances the effective mass to nearly one thousand times the bare electron mass, in this heavy fermion metal which does not have high energy strong-coupled phonons.  Recent dynamic mean field theory calculations however, have shown that robust $s$-wave superconductivity driven by \textit{local} spin-fluctuations is found in solutions to the Kondo-lattice model which is commonly used to describe heavy-fermion metals \cite{Bosensiek}.  Other recent theoretical work has shown that electron-phonon coupling could be strongly enhanced near a QCP again stabilizing $s$-wave superconductivity \cite{Tazai}.  Our results might therefore support a new type of unconventional superconductivity where the gap function is $s$-wave but the pairing is nevertheless driven by strong magnetic fluctuations.}

\section*{Materials and methods}

$S$-type single crystals of CeCu$_2$Si$_2$ were grown by the flux method~\cite{Seiro}. Specific heat was measured by the standard quasi-adiabatic heat-pulse method. The temperature dependence of penetration depth $\lambda(T)$ was measured by using the tunnel diode oscillator technique operating at $\sim14$\,MHz. Weak ac magnetic field ($\sim 1\,\mu$T) is applied along the $c$ axis inducing screening currents in the $ab$ plane. Thermal conductivity was measured by the standard steady state method using one heater and two thermometers, with an applied temperature gradient less than 2\,\% of the sample temperature. The contacts were made by indium solder with contact resistance much less than 0.1\,$\Omega$. We examined the effect of superconductivity of indium by applying small magnetic field and found no discernible difference. We also measured with contacts made of silver paint with contact resistance less than 0.1\,$\Omega$ and observed identical results. We measured several different crystals and obtained essentially the same results.

Electron irradiation was performed in the electron irradiation facility SIRIUS at {\'E}cole Polytechnique. We used electrons with incident energy of 2.5\,MeV for which the energy transfer from the impinging electron to the lattice is above the threshold energy for the formation of vacancy interstitial (Frenkel) pairs that act as point defects \cite{Mizukami}. In order to prevent the point defect clustering, irradiation is performed at 25\,K using a H$_2$ recondenser.  {According to the standard calculations, the penetration range for irradiated electrons with 2.5\,MeV energy is as long as $\sim2.75$\,mm, which is much longer than the thickness of the crystals (typically 50-100\,$\mu$m). This endures that the point defects created by the irradiation are uniformly distributed throughout the sample thickness. Our simulations also show that} for 1\,C/cm$^2$ dose, irradiation causes about 1-2 vacancies per 1000 Ce atoms.  However, because of the defect annihilations due to annealing effect at room temperature, the number of vacancies is not directly proportional to the dose density.


\vspace{1cm}

{\bf Acknowledgements: }
We thank  A. Chubukov, P. Coleman, P. Hirschfeld, K. Ishida, H. Kontani, H. v. L\"{o}hneysen, P. Thalmeier, C. Varma, I. Vekhter, and Y. Yanase   for useful discussions.
{\bf Funding: }
This work was supported by Grants-in-Aid for Scientific Research (KAKENHI) (No.\,25220710, No.\,15H02106) and Grants-in-Aid for Scientific Research on Innovative Areas ``Topological Materials Science" (No.\,15H05852) and ``J-Physics'' (No.\,15H05883) from Japan Society for the Promotion of Science (JSPS),
and by the UK Engineering and Physical Sciences Research Council (grant no.\,EP/L025736/1 and EP/L015544/1).
{\bf Author Contributions: }
Y.Matsuda and T.Shibauchi  conceived and designed the study. T.Y., Y.T. Y.K. and D.T. performed the thermal conductivity measurements. T.T., J.A.W., Y.Mizukami, C.P., A.C. and T.Shibauchi carried out penetration depth measurements. T.Y., T.T., Y.T., Y.Mizukami and D.T carried out electrical resistivity measurements. Y.T. S.K. and T.Sakakibara performed specific heat measurements. M.K. and Y. Mizukami electron-irradiated the samples. H.S.J., S.S. and C.G. synthesized the high-quality single crystalline samples. H.I. performed calculations based on the first-principles calculations. T.Shibauchi, Y.T.,  A.C. and Y.Matsuda discussed and interpreted the results and prepared the manuscript.
{\bf Competing interests: }
The authors declare that they have no competing interests.
{\bf Data and materials availability:} All data needed to evaluate the conclusions in the paper are present in the paper and/or the Supplementary Materials. Additional data related to this paper may be requested from the authors.

\newpage
\renewcommand{\figurename}{FIG. S$\!\!$}
\renewcommand{\tablename}{Table S$\!\!$}
\renewcommand{\theequation}{S\arabic{equation}}
\newcommand{\beq}{\begin{equation}}
	\newcommand{\eeq}{\end{equation}}
\newcommand{\ud}{\mathrm{d}}

\newcolumntype{Y}{>{\centering\arraybackslash}X}

\section*{SUPPLEMENTARY MATERIALS}
\setcounter{figure}{0}

\subsection*{section SI. Temperature dependence of penetration depth}
The data for the temperature dependence of the penetration depth $\Delta\lambda(T)$ measured in two samples show consistent results (Fig.\,S1), which can be fitted to the exponential dependence expected for the fully gapped superconductors. 

In Fig.\,S2, the $T$-exponent $n$ of the $\Delta\lambda(T)$ data in the power-law fit and the effective gap $\Delta_e$ in the exponential fit are plotted as a function of the upper limit of the fit, $T_{\rm max}$. In clean and dirty $d$-wave superconductors (or more generally superconductors with line nodes) power-law dependencies with exponents 1 and 2 are expected (Fig.\,S2A, dashed lines). In a multigap system the effective gap $\Delta_e$ is close to the minimum one on all sheets of Fermi surface.

\begin{figure}[!b]
	\begin{center}
		\includegraphics[width=0.7\linewidth]{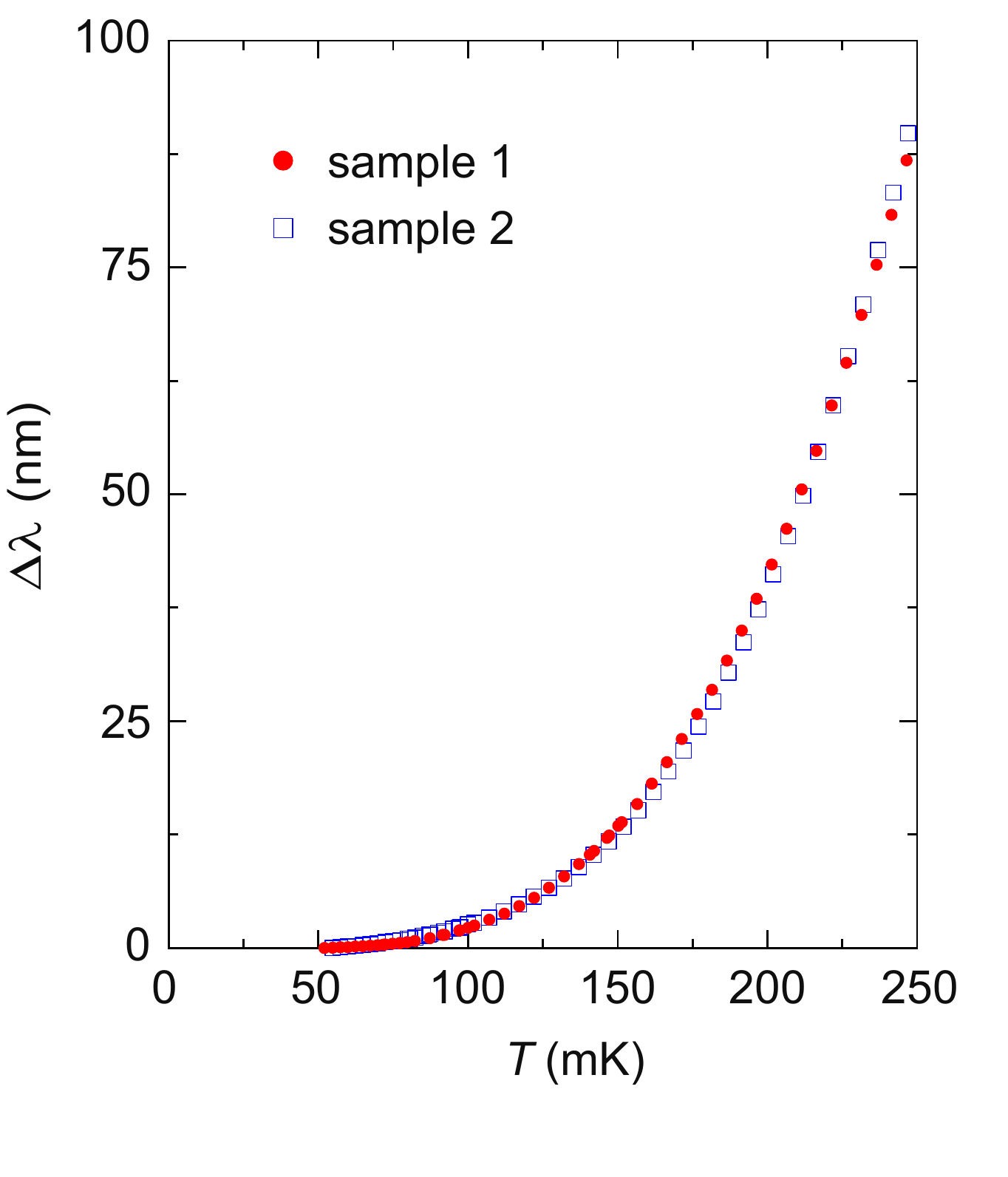}
	\end{center}
	\vspace*{-10mm}
	\caption{{\bf Magnetic penetration depth versus temperature for two samples measured.} Sample 1 is the same as show in Fig.\,3B. The data for sample 2 have been multiplied by 1.16 for comparison to sample 1. The temperature dependence is almost identical but the scale factors differ by 20\% which is within the bound of the expected error on the geometrical scale factors.}
\end{figure}

\begin{figure}[!t]
	\begin{center}
		\includegraphics[width=\linewidth]{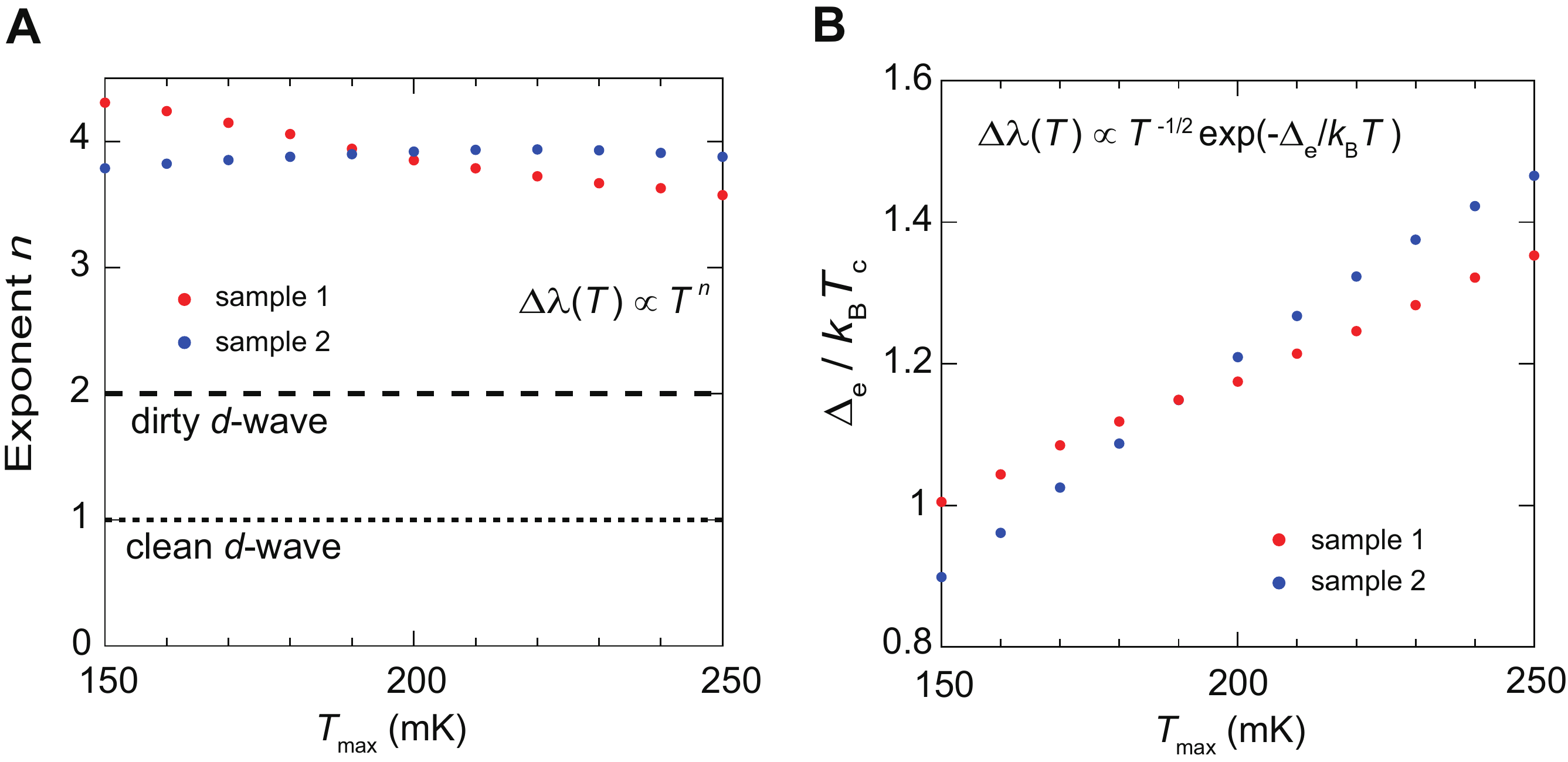}
	\end{center}
	\vspace*{-5mm}
	\caption{{\bf Parameters obtained for the fits to the $\boldsymbol{\Delta\lambda(T)}$ data.} ({\bf A} and {\bf B}) The low-temperature penetration depth data are fitted by using the power-law $T^n$ ({\bf A}) and exponential temperature dependence $T^{-1/2}\exp(-\Delta_e/k_BT)$ ({\bf B}), where the fitting temperature range is chosen up to the maximum temperature $T_{max}$. }
\end{figure}

\subsection*{section SII. Lower critical field}
Our radio-frequency inductive measurements measure very precisely the temperature dependence of the magnetic penetration depth $\lambda$ relative to some reference level at low temperature, but not its absolute value. To determine the absolute value we have performed measurements of the lower critical field $H_{c1}$ using a micro-Hall probe array.  An array of Hall sensors is placed below the sample and so the magnetic induction $B$ at the position of the Hall sensor is measured as a function of the applied field $H$. Figure\,S3 shows a $B$($H$) curve for a Hall sensor close to the center of a sample of CeCu$_2$Si$_2$ of approximate dimensions 0.29$\times$0.40$\times$0.09\,mm$^3$. At each temperature the sample was cooled in nominally zero field and the field increased towards the maximum.  The sample was then warmed above $T_c$ and cooled again in zero field and the field increased towards the negative limit. The magnetic field was produced by a copper solenoid so the remnant field from this was very low but the earth's field was not shielded. Tests showed that cooling in a small (positive or negative) field of less than 0.1\,mT did not change the results.

\begin{figure}[b]
	\begin{center}
		\includegraphics[width=\linewidth]{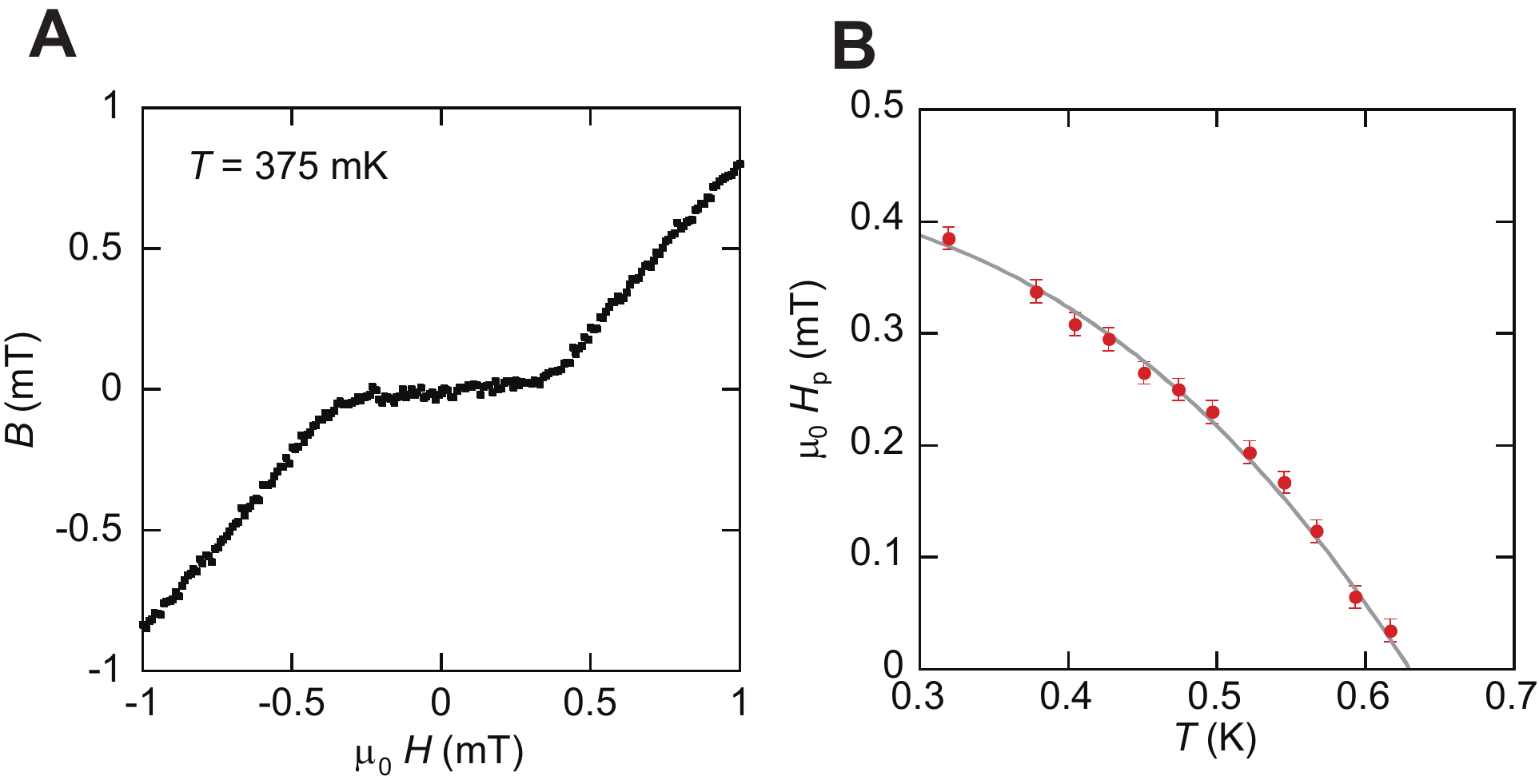}
	\end{center}
	\vspace*{-5mm}
	\caption{{\bf  Lower critical field measurements of CeCu$\boldsymbol{_2}$Si$\boldsymbol{_2}$.} ({\bf A}) Magnetic induction ($B$) versus applied field ($H$) measured by a Hall sensor a few microns below the middle of the sample. Two sweeps are shown, one for increasing field and another for decreasing field. The sample was zero field cooled before each sweep. ({\bf B})  Field of first flux penetration $H_p$ versus temperature. The line is a guide to the eye.}
\end{figure}

At low field there is a small linear increase in $B$ with $H$ because of incomplete flux shielding of the sensor by the sample which sits a few microns above. At a well-defined field $H_p$ flux enters the sample and $B$ increases rapidly with increasing $H$. Taking the average of $H_p$ for the positive and negative field sweeps cancels out the offset due to the earth's field. We relate $H_p$ to $H_{c1}$ using the following relation from Brandt [43] for a strip

\beq
\frac{H_p}{H_{c1}} = {\rm tanh} \sqrt{\frac{0.36 c}{a}} \nonumber
\eeq
where $a$ is the shorter of the in-plane dimensions of the sample and $c$ is the $c$-axis dimension (our measurements are performed with $B\parallel c$).  At $T=320$\,mK we find $H_p=0.39$\,mT and $H_{c1}=1.2\pm0.1$\,mT, where the error includes uncertainty in the sample dimensions. Then solving the Ginzburg-Landau equation

\beq
\mu_0H_{c1}=\frac{\phi_0}{4\pi\lambda^2}\left[\ln\left(\frac{\lambda}{\xi}\right)+0.5\right]\nonumber
\eeq
with $\xi=4.7$\,nm, gives $\lambda$($T=320$\,mK)\,$=890\pm40$\,nm. Then we use the radio frequency measured change in $\lambda$ from 50\,mK to 320\,mK, $\Delta\lambda=190\pm25$\,nm, to calculate $\lambda(T=0)=700\pm50$\,nm. Repeating this procedure for a second sample, with dimensions $0.35\times0.32\times0.041$\,mm$^3$, gave a consistent result within the error, $\lambda(T=0)=680\pm50$\,nm.

\subsection*{section SIII. Zero-field thermal conductivity}

The thermal conductivity in the superconducting state can be written as a sum of the quasiparticle and phonon contributions, $\kappa=\kappa_{\rm qp}+\kappa_{\rm ph}$.   The phonon conductivity in the boundary-limited scattering regime at low temperature is expressed as,  $\kappa_{\rm ph} = \frac{1}{3} \beta \langle v_{s} \rangle \ell_{\rm ph} T^{3}$,  where $\beta$ is the phonon specific heat coefficient, $\langle v_{s} \rangle$ is the mean acoustic phonon velocity, and $\ell_{\rm ph}$ is the phonon mean free path.  At low temperatures,  $\kappa_{\rm ph}$ shows a power-law dependence on temperature; $\kappa_{\rm ph}\propto T^{\alpha}$ and $\alpha$ ranges from 2 to 3, depending on the nature of the surface scattering.   In real systems, $\alpha$ takes a value intermediate between 2 and 3 [44].  The best fit in a wide $T$-range is obtained for $\alpha=2.4$ for $\kappa_{a}$.  In this case, the residual term is close to zero (Fig.\,S4B).

\begin{figure}[t]
	\begin{center}
		\includegraphics[width=\linewidth]{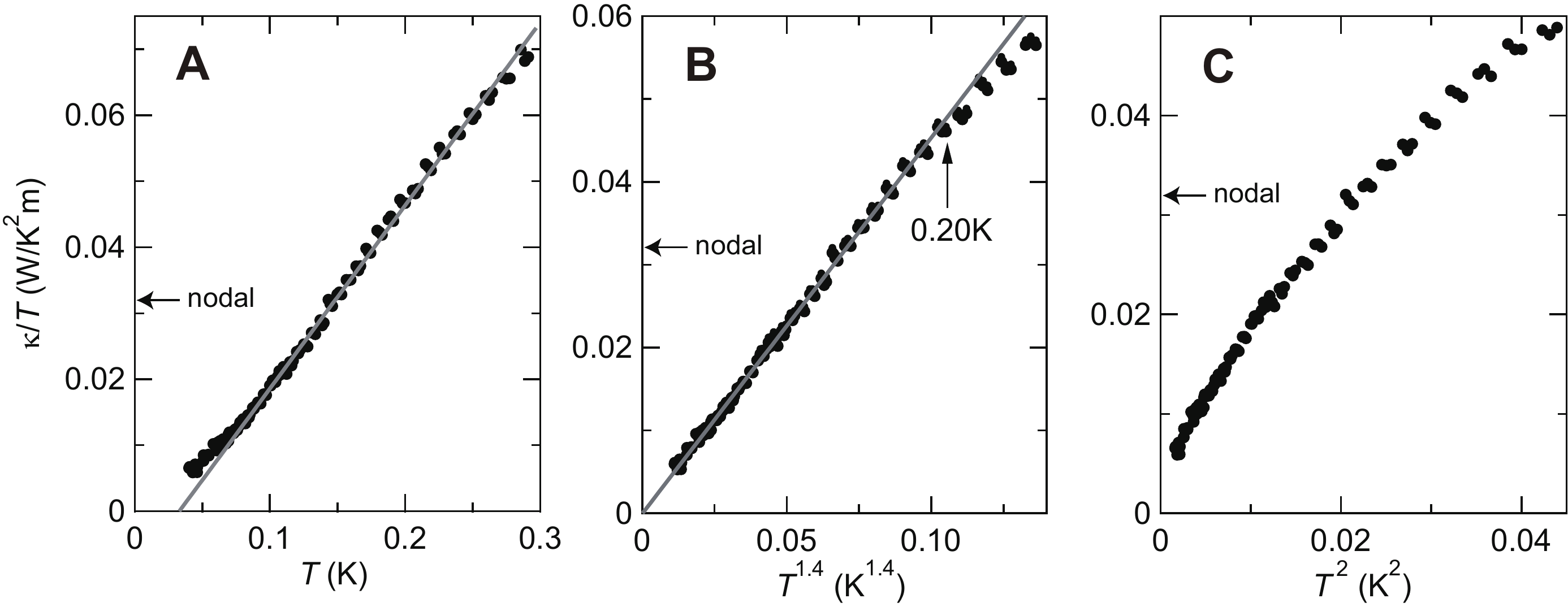}
	\end{center}
	\vspace*{-5mm}
	\caption{{\bf Temperature dependence of thermal conductivity at low temperatures.} ({\bf A}) Thermal conductivity divided by temperature $\kappa/T$ of CeCu$_2$Si$_2$ plotted against $T$ for heat current along the $a$-axis in zero field. ({\bf B}) The same data of $\kappa/T$ plotted against $T^{1.4}$. ({\bf C}) The same data plotted against $T^2$. Solid lines are linear fits to the data. The arrow in the middle panel marks the temperature where the fits starts to deviate from the linear behavior.  The value of $\kappa/T$ expected for a line nodes is indicated.}
\end{figure}

For comparison, we show $\kappa/T$ plotted against $T$, $T^{1.4}$ and $T^2$ in Fig.\,S4. When $\kappa/T$ is plotted against $T$, a linear extrapolation to zero temperature results in an unphysical negative value. Moreover, $\kappa/T$ deviates from the fit at the lowest temperature. When $\kappa/T$ is plotted against $T^{1.4}$ as shown in the middle panel, it is linearly fitted up to 0.2\,K without deviation at low temperature. The right panel shows $\kappa/T$ plotted against $T^2$. The data exhibit a convex curvature, indicating that $\kappa/T$ does not follow a $T^2$ dependence.

In a superconductor with line nodes, a finite residual thermal conductivity $\kappa_0/T\equiv \kappa/T (T\rightarrow 0)$ is expected due to the existence of a residual normal fluid, which is a consequence of impurity scattering, even for low concentrations of non-magnetic impurities [45].   At $T=40$\,mK, in-plane phonon conductivity $\kappa_a^{\rm ph}/T$ estimated by using the Wiedemann-Franz law in the normal state, $\kappa_a^{\rm ph}/T=\kappa_a(H_{c2})/T-L_0/\rho_a(T)$, is  $\sim 4$\,mW/K$^2$m, which yields  in-plane quasiparticle thermal conductivity $\kappa_a^{\rm qp}/T\sim $3\,mW/K$^2$m in zero field. The residual  thermal conductivity expected for line node  is estimated as $\kappa_{a0}/T\approx (L_0/\rho_{a0})\cdot(\xi_{ab}/\ell_{ab}) \sim$ 32\,mW/K$^2$m [45].  Here $\xi_{ab}=4.7$\,nm is the in-plane coherence length estimated by the orbital limited upper critical field of 14.7\,T for {\boldmath $H$}$\parallel c$ [20] 
and $\ell_{ab}\sim 8$\,nm is the in-plane mean free path obtained from $\ell_{ab}=v_F^{ab}\lambda^2(0)\mu_0/\rho_{a0}$, using $\rho_{a0}=43\,\mu\Omega$cm and the average of in-plane Fermi velocity $v_F^{ab}\sim 5800$\,m/s for the hole band calculated by LDA+U, taking into account the mass renormalization $z=1/50$ which is determined by the specific heat measurements.    These results indicate that the residual thermal conductivity at $T \rightarrow 0$, if present, is considerably smaller than that expected for line node (Fig.\,S4).  A similar conclusion is obtained for the out-of-plane thermal conductivity $\kappa_c$ ($\bm{Q}\parallel c$).

\subsection*{section SIV. Impurity effect of superconductivity}

In Fig.\,S5, the impurity effect of superconductivity in CeCu$_2$Si$_2$ is compared with those of CeCoIn$_5$ [27], YBa$_2$Cu$_3$O$_{7-\delta}$ [28], Ba(Fe$_{0.76}$Ru$_{0.24}$)$_2$As$_2$ [46], and the Abrikosov-Gor'kov (AG) theory for an isotropic $s$-wave superconductor with magnetic impurities. As seen clearly, the suppression of $T_c$ is much weaker than these compared superconductors with sign change in the gap function. We also plot the data for the $s$-wave superconductors MgB$_2$ [30] and YNi$_2$B$_2$C [31] with very anisotropic energy gaps.

Here we note that it has been suggested that in iron-based superconductors with $s_\pm$ symmetry, the superconductivity may be more robust against the impurity than $d$-wave superconductors.  This is because in pnictude the momentum of the intraband scattering is much smaller than that of interband scattering owing to their characteristic Fermi surface with well separated very small electron and hole pockets.  In such a case, the interband scattering, which is responsible for the pairing interaction, is less effective  against the impurity than the intraband scattering. 

\begin{figure}[t]
	\begin{center}
		\includegraphics[width=\linewidth]{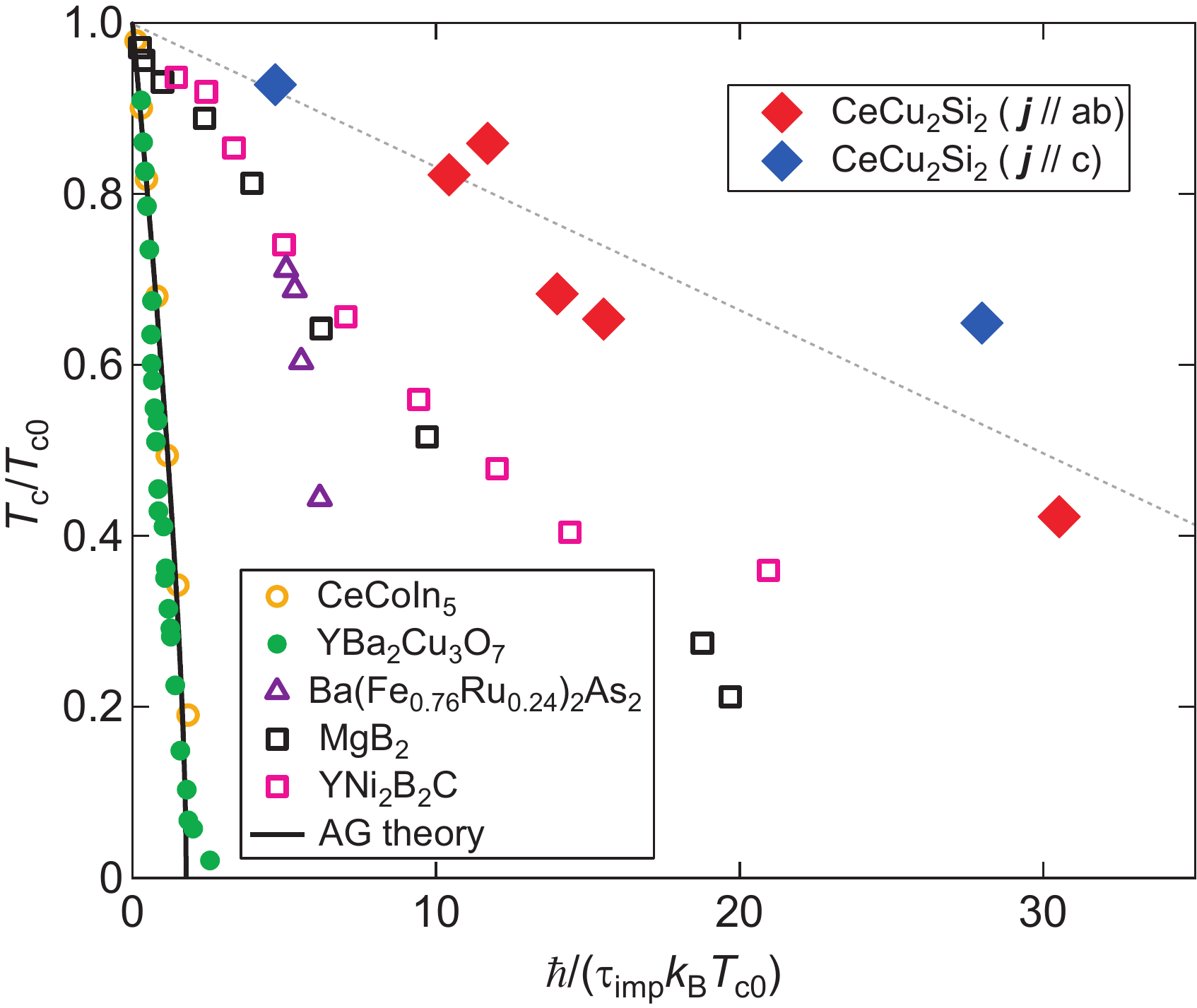}
	\end{center}
	\vspace*{-5mm}
	\caption{{\bf Comparison of impurity effect of CeCu$\boldsymbol{_2}$Si$\boldsymbol{_2}$ with those of other superconductors.}  Suppression of superconducting transition temperature $T_c/T_{c0}$ as a function of dimensionless scattering rate $\hbar/\tau_{\rm imp}k_BT_{c0}$, where $\tau_{\rm imp}$ is the impurity scattering time estimated from residual resistivity $\rho_0$ and the penetration depth, $\tau_{\rm imp}=\mu_0\lambda_{ab}\lambda_c/\rho_0$. The solid line shows the prediction of the Abrikosov-Gorkov (AG) theory for an isotropic $s$-wave superconductor with magnetic impurities. We also plot the data for Sn-substituted CeCoIn$_5$ ($d$-wave) [27], 
		electron-irradiated YBa$_2$Cu$_3$O$_{7-\delta}$ ($d$-wave) [28], 
		electron-irradiated Ba(Fe$_{0.76}$Ru$_{0.24}$)$_2$As$_2$ (possibly $s_{\pm}$-wave) [46],  and neutron-irradiated MgB$_2$ [30] and YNi$_2$B$_2$C [31]. 
		The value of $T_{c0}$ is estimated by extrapolating two initial data points to zero $1/\tau_{\rm imp}$ limit. Rather weak pair-breaking effect in Ba(Fe$_{0.76}$Ru$_{0.24}$)$_2$As$_2$ has been attributed to a large imbalance between intra- and inter-band scattering [46]. 
		For MgB$_2$ data, we use the value of $\lambda_{ab}(0)=\lambda_c(0)=100$\,nm  [47].  For  YNi$_2$B$_2$C data, we use $\lambda_{ab}(0)=110$\,nm [48] and  $\lambda_{c}(0)=\lambda_{ab}(0)H_{c2}^{ab}/H_{c2}^{c}=140$\,nm [49], where $H_{c2}^{ab}$ and $H_{c2}^c$ are upper critical field parallel and perpendicular to the $ab$ plane. 
	}
\end{figure}

\end{document}